\documentclass{elsart}
\usepackage{graphicx}
\usepackage{dcolumn}
\usepackage{bm}
\usepackage{amsmath}
\usepackage{amsfonts}
\usepackage{amssymb}
\usepackage{subfigure}
\providecommand{\U}[1]{\protect\rule{.1in}{.1in}}

\begin{document}
\begin{frontmatter}
\title{Quantum walk, entanglement and thermodynamic laws}
\author{Alejandro Romanelli},
\address{Instituto de F\'{\i}sica, Facultad de Ingenier\'{\i}a\\
Universidad de la Rep\'ublica\\ C.C. 30, C.P. 11000, Montevideo, Uruguay}

\date{\today}

\begin{abstract}
We consider an special dynamics of a quantum walk (QW) on a line. Initially, the walker localized at the origin of the
line with arbitrary chirality, evolves to an asymptotic stationary state. In this stationary state a measurement is
performed and the state resulting from this measurement is used to start a second QW evolution to achieve  a second
asymptotic stationary state. In previous works, we developed the thermodynamics associated with the entanglement between
the coin and position degrees of freedom in the QW. Here we study the application of the first and second laws of
thermodynamics to the process between the two stationary states mentioned above. We show that: i) the entropy change has
upper and lower bounds that are obtained analytically as a function of the initial conditions. ii) the energy change is
associated to a heat-transfer process.
\end{abstract}

\end{frontmatter}

\section{Introduction}
Quantum walks (QWs) constitute the quantum analogue of classical
random walks \cite{Aharonov} and also the quantum version of
cellular automata \cite{Meyer}. They have been intensively
investigated, especially in connection with quantum information
science \cite{Watrous,Nayak,Ambainis,Kempe,Kendo1,Kendo2,Konno}. As
in the classical case, QWs have been proposed as elements to design
quantum algorithms \cite{Shenvi,Ambainis2,Childs0,Childs1} and more
recently it has been shown that they can be used as a universal
model for quantum computation \cite{Childs,Lovett}.

We have been investigating \cite{alejo2010,alejo2011,armando2011} the asymptotic behavior of the QW on a line, focusing
on the probability distribution of chirality independently of position. We showed that this distribution has a stationary
long-time limit that depends on the initial conditions and that it is possible to define a thermodynamic equilibrium
between the degrees of freedom of position and chirality \cite{Alejo2012,gustavo2014,renato2014,JCmodel}. For this
equilibrium state we have introduced a temperature concept for an unitary closed system.

On the other hand, the fundamental lower bound of the thermodynamic
energy cost of information processing has been a topic of active
research \cite{Sagawa1,Sagawa2}. On average, the minimum amount of
work required to erase $1$ bit of information from a memory is
$\kappa T \ln2$ \cite{Landauer}. In the last decades developments
in nano-science have enabled the direct measurement of such
minuscule amounts of work for small non-equilibrium thermodynamic
systems \cite{nano}. At the same time recent advances in
technology have opened the possibility of building useful quantum
computing devices \cite{Cohen}. Therefore, it seems essential to
identify bounds on the thermodynamic energy cost of information
processing \cite{chuang} for these new quantum devices.

In the present paper we study the relationship between the QW
thermodynamics and information processing. In particular, we show
that it is possible to apply the thermodynamic laws  to the QW
dynamics after a measurement process. We obtain the upper and lower
bound for the asymptotic change of the entanglement entropy. Our
result may be thought as complementary to the results presented in
Refs. \cite{Sagawa1,Sagawa2} where the information content and
thermodynamic variables are treated on an equal footing.

The paper is organized as follows. In the next section the usual QW on a line is presented, in the third section the
system dynamics with measurement is developed, in the fourth section the entropy change between the asymptotic stationary
states is studied, in the fifth section the laws of thermodynamics are applied to the same process. Finally, in the last
section we draw some conclusions.

\section{QW on a line}

The composite Hilbert space of the QW is the tensor product $\mathcal{H}_{T}\otimes \mathcal{H}_{\pm}$ where
$\mathcal{H}_{T}$ is the Hilbert space associated to the motion on a line and $\mathcal{H}_{\pm}$ is the chirality (or
coin) Hilbert space. In this composite space the walker moves, at discrete time steps $t\in \mathbb{N}$, along a
one-dimensional lattice of sites $k\in \mathbb{Z}$. The direction of motion depends on the chirality states, either right
or left. The wave vector can be expressed as the spinor
\begin{equation}
|\Psi (t)\rangle =\sum\limits_{k=-\infty }^{\infty }\left(
\begin{array}{c}
a_{k}(t) \\
b_{k}(t)%
\end{array}%
\right) |k\rangle ,  \label{spinor}
\end{equation}%
where the upper (lower) component is associated to the left (right) chirality. The QW is ruled by a unitary map whose standard form is \cite%
{Romanelli09,Alejo2,Alejo1,Alejo4}
\begin{align}
a_{k}(t+1)& =a_{k+1}(t)\,\cos \theta \,+b_{k+1}(t)\,\sin \theta ,\,
\notag
\\
b_{k}(t+1)& =a_{k-1}(t)\,\sin \theta \,-b_{k-1}(t)\,\cos \theta ,\label{mapa}
\end{align}%
where $\theta \in \left[ 0,\pi /2\right] $ is a parameter defining
the bias of the coin toss. Here we take $\theta =\frac{\pi }{4}$ for
an unbiased or Hadamard coin. The probability of finding the walker
at $\left( k,t\right) $ is
\begin{equation}
P(k,t)=\left\vert a_{k}(t)\right\vert ^{2}+\left\vert b_{k}(t)\right\vert ^{2}. \label{prob}
\end{equation}%
The global left and right chirality probabilities are defined as
\begin{align}
P_{L}(t)& \equiv \sum_{k=-\infty}^{\infty }\left\vert a_{k}(t)\right\vert ^{2},\,  \notag \\
P_{R}(t)& \equiv \sum_{k=-\infty }^{\infty }\left\vert
b_{k}(t)\right\vert ^{2},  \label{chirality}
\end{align}%
with $P_{R}(t)+P_{L}(t)=1$ and the interference term is defined as
\begin{equation}
Q(t)\equiv \sum_{k=-\infty }^{\infty }a_{k}(t)b_{k}^{\ast }(t).
\label{qdet}
\end{equation}
In the generic case $Q(t)$ together with $P_{L}(t) $ and $P_{R}(t)$
are time depend functions that have long-time limiting values
\cite{alejo2010} which are determined both by the initial conditions
and by the map in Eq.(\ref{mapa}). The relation between the initial condition and the asymptotic
distributions has also been recently
explored in Ref.\cite{Montero}. Let us call the mentioned limits as
\begin{align}
\Pi _{L}& \equiv
\begin{array}{c}
\lim \text{ }P_{L}(t) \\
t\rightarrow \infty~~~~%
\end{array}%
,\,  \notag \\
\Pi _{R}& \equiv
\begin{array}{c}
\lim \text{ }P_{R}(t) \\
t\rightarrow \infty~~~~%
\end{array}%
,\,  \notag \\
Q_{0}& \equiv
\begin{array}{c}
\lim \text{ }Q(t) \\
t\rightarrow \infty~~%
\end{array}%
=\mu+i\nu,\,  \label{asym}
\end{align}%
where $\mu$ and $\nu$ are respectively the real and imaginary part
of $Q_0$. The following relations are verified \cite{alejo2010}
between $\Pi _{L}$, $\Pi _{R}$ and $Q_{0}$
\begin{align}
\Pi _{L}& \equiv \frac{1}{2}+\mu ,\,  \notag \\
\Pi _{R}& \equiv \frac{1}{2}-\mu . \label{estacio}
\end{align}
It is important to emphasize that the asymptotic behavior in Eq.(\ref{asym}) is determined by the interference term
$Q_{0}$ that only depends on the initial conditions.

The initial condition the walker localized at the origin with arbitrary chirality will play a central roll in our
analytic treatment. Then
\begin{equation}
|\Psi_1 (0)\rangle =\left(
\begin{array}{c}
\cos ({\gamma }/{2}) \\
\exp i\varphi \text{ }\sin ({\gamma }/{2})
\end{array}
\right) |0\rangle ,  \label{psi0}
\end{equation}
where $\gamma \in \left[ 0,\pi \right] $ and $\varphi \in \left[ 0,2\pi\right]$ define a point on the unit Bloch sphere.
We obtain \cite{alejo2010}
\begin{equation}
Q_{0}=\frac{1}{2}(1-\frac{1}{\sqrt{2}})\left[ \cos \gamma \text{
}+\sin \gamma \text{ }(\cos \varphi +i\sqrt{2}\sin \varphi )\right]
, \label{q0entengl}
\end{equation}
and also $\Pi _{L}$ and $\Pi _{R}$ using Eq.(\ref{estacio}).

\section{Dynamical evolution and measurement}
\subsection{First step} We consider first the QW evolution starting from the initial
condition given by Eq.(\ref{psi0}) and determine the asymptotic
density matrix. The quantum density matrix is defined as
\begin{equation}
\rho (t)=|\Psi (t)\rangle \langle \Psi (t)|,  \label{density}
\end{equation}%
substituting Eq.(\ref{spinor}) into Eq.(\ref{density}) we have
\begin{equation}
\rho (t)=\sum_{k,k^{\prime }}\left(
\begin{array}{cc}
a_{k}(t)a_{k^{\prime }}^{\ast }(t) & a_{k}(t)b_{k^{\prime }}^{\ast }(t) \\
a_{k}^{\ast }(t)b_{k^{\prime }}(t) & b_{k}(t)b_{k^{\prime }}^{\ast }(t)%
\end{array}%
\right) |k\rangle \langle k^{\prime }|,  \label{density1}
\end{equation}%
where $a_{k}(t)$ and $b_{k}(t)$ depend also on the initial
conditions $\gamma $ and $\varphi $.

Note that when
$t\rightarrow\infty$ the limits of $a_{k}(t)$ and $b_{k}(t)$ are not
 defined because, in general, they have an oscillatory asymptotic behavior \cite{Nayak}; however the
limits given by Eq.(\ref{asym}) are always well defined. In the following we call $\mathbf{a}_{k}$ and  $\mathbf{b}_{k}$
to the values of $a_{k}(t)$ and $b_{k}(t)$ respectively, evaluated at times large enough so that the asymptotic limit,
Eq.(\ref{asym}), is essentially attained.

Let us define the asymptotic reduced density matrix as $\rho _{1c}=\lim\mathrm{tr}(\rho )=\lim\sum_{l}\langle
l|\rho|l\rangle$ for $t\rightarrow\infty$. This matrix takes the following shape
\begin{equation}
\rho_{1c}=\left(
\begin{array}{cc}
\Pi _{L}& Q_{0}\\
{Q}^{\ast }_{0} &\Pi _{R}
\end{array}%
\right). \label{densityc}
\end{equation}
\subsection{Second step}
When the asymptotic density matrix is attained a measurement of position and chirality is performed. Then the wave
function collapses into one element of the set of eigenvectors of the measurement operator $\{|k\rangle |\pm \rangle\}$,
where
\begin{align}
|+\rangle & =\left(
\begin{array}{c}
1 \\
0%
\end{array}%
\right) ,\,  \notag \\
|-\rangle & =\left(
\begin{array}{c}
0 \\
1%
\end{array}%
\right) .  \label{e1e2}
\end{align}%
After the measurement, the density matrix has the form
\begin{equation}
\rho _{1}=\sum_{k=-\infty }^{\infty }|\mathbf{a}_{k}|^{2}|k+\rangle
\langle k+|+\sum_{k=-\infty }^{\infty
}|\mathbf{b}_{k}|^{2}|k-\rangle \langle k-|,  \label{density3}
\end{equation}%
where $|k{\pm }\rangle =|k\rangle |{\pm }\rangle $. Eq.(\ref{density3}) can be written in matrix form as
\begin{equation}
\rho _{1}\equiv\sum_{k}\left(
\begin{array}{cc}
|\mathbf{a}_{k}|^{2} & 0 \\
0 & |\mathbf{b}_{k}|^{2}%
\end{array}%
\right) |k\rangle \langle k|.  \label{density2}
\end{equation}%
Equation (\ref{density3}) sets a new initial condition and we
will study the asymptotic evolution of the new density matrix.
Essentially we must study the evolution
of the densities $|k \pm \rangle \langle k\pm |$. We know, from the QW
dynamics, that if the initial condition is given by Eq.(\ref{psi0})
then the density matrix is given by Eq.(\ref{density1}). We can use
these equations as a recipe to obtain the unknown evolutions, that
is, when we choose a particular initial condition given by $|k \pm
\rangle$ we have
\begin{align}
& U(t)|k \pm \rangle \langle k \pm|U^{\dag}(t)=
\notag \\
& \sum_{n,n^{\prime }}\left(
\begin{array}{cc}
 {a}_{n\pm}(t,k) {a}_{n^{\prime }\pm}^{\ast }(t,k) &  {a}_{n\pm}(t,k) {b}_{n^{\prime}\pm }^{\ast }(t,k) \\
 {a}_{{n}^{\ast }\pm }(t,k) {b}_{n^{\prime }\pm}(t,k) &  {b}_{n\pm }(t,k) {b}_{n^{\prime}\pm
}^{\ast }(t,k)
\end{array}%
\right) |n\rangle \langle n^{\prime }|,  \label{density111}
\end{align}%
where $U(t)$ is the QW evolution operator and $ {a}_{n\pm}(t,k)$ and $  {b}_{n\pm }(t,k)$ verify the map Eq.(\ref{mapa}),
that is $ {a}_{n\pm}(t,k)$ and $ {b}_{n\pm}(t,k)$ are equivalent to some functions $a_{n}(t)$ and $b_{n}(t)$. The
expressions $ {a}_{n\pm}(t,k)$ and $  {b}_{n\pm }(t,k)$ show explicitly their dependence both on the chirality,
$\{|+\rangle,|-\rangle\}$, as with, $k$, the walker's initial position on a line. Therefore, the new quantum density
matrix is
\begin{align}
\rho (t)&=U(t)\rho _{1}U^{\dag}(t)
\notag \\
& =\sum_{n,n^{\prime }}\left(
\begin{array}{cc}
A_{11}(n,n^{\prime })&A_{12}(n,n^{\prime }) \\
A_{21}(n,n^{\prime })&A_{22}(n,n^{\prime })
\end{array}
\right) |n\rangle \langle n^{\prime }| ,  \label{density30}
\end{align}
where
\begin{align}
A_{11}(n,n^{\prime })=&\sum_{k}\left[|\mathbf{a}_{k}|^{2} {a}_{n+}(t,k) {a}_{n^{\prime }+}^{\ast }(t,k)\right.  \notag \\
+&\left.|\mathbf{b}_{k}|^{2}  {a}_{n-}(t,k) {a}_{n^{\prime }-}^{\ast
}(t,k)\right], \label{density6}
\end{align}
\begin{align}
A_{22}(n,n^{\prime })=&\sum_{k}\left[|\mathbf{a}_{k}|^{2}
 {b}_{n+}(t,k) {b}_{n^{\prime }+}^{\ast }(t,k)\right.  \notag \\
+&\left.|\mathbf{b}_{k}|^{2}  {b}_{n-}(t,k) {b}_{n^{\prime }-}^{\ast
}(t,k)\right], \label{density9}
\end{align}
\begin{align}
A_{12}(n,n^{\prime })=&\sum_{k }\left[|\mathbf{a}_{k}|^{2}
 {a}_{n+}(t,k) {b}_{n^{\prime }+}^{\ast }(t,k)\right.  \notag \\
+&\left.|\mathbf{b}_{k}|^{2}  {a}_{n-}(t,k) {b}_{n^{\prime }-}^{\ast
}(t,k)\right], \label{density7}
\end{align}
\begin{align}
A_{21}(n,n^{\prime })=&A_{12}^{\ast }(n,n^{\prime }). \label{density8}
\end{align}

We again point out that the probability density has an asymptotic
limit for long times. We now call  the values of ${a}_{n\pm}(t,k)$
and $ {b}_{n\pm}(t,k)$ evaluated at such times
$\mathbf{{a}}_{n\pm}(k)$ and $\mathbf{{b}}_{n\pm}(k)$ respectively.
We want to calculate in this limit the reduced density matrix,
namely $\rho _{2c}=\lim\mathrm{tr}(\rho )=\lim\sum_{l}\langle
l|\rho|l\rangle$, $t\rightarrow\infty$. Using the density matrix
given by Eq.(\ref{density30}), we have
\begin{equation}
\rho_{2c}=\begin{array}{c} \lim\sum_{l=-\infty}^{\infty}\left(
\begin{array}{cc}
A_{11}(l,l)&A_{12}(l,l) \\
A_{12}^{\ast }(l,l)&A_{22}(l,l)
\end{array}
\right), \\
t\rightarrow\infty ~~~~~~~~~~~~~~~~~~~~~~~~~~~~~~~~~~~~~
\end{array}
\label{density300}
\end{equation}
where
\begin{align}
\begin{array}{c}
  \lim\sum_{l=-\infty}^{\infty}A_{11}(l,l) \\
  t\rightarrow\infty~~~~~~~~~~~~~~~~~~
\end{array}
=&\sum_{k}|\mathbf{a}_{k}|^{2}\sum_{l}| \mathbf{{a}}_{l+}(k)|^{2}  \notag \\
+&\sum_{k}|\mathbf{b}_{k}|^{2} \sum_{l}| \mathbf{{a}}_{l-}(k)|^{2},
\label{ty6}
\end{align}
\begin{align}
\begin{array}{c}
  \lim\sum_{l=-\infty}^{\infty}A_{12}(l,l) \\
  t\rightarrow\infty~~~~~~~~~~~~~~~~~~
\end{array}
=&\sum_{k}|\mathbf{a}_{k}|^{2}\sum_{l} \mathbf{{a}}_{l+}(k) \mathbf{{b}}_{l+}^{\ast }(k)  \notag \\
+&\sum_{k}|\mathbf{b}_{k}|^{2} \sum_{l} \mathbf{{a}}_{l-}(k)
\mathbf{{b}}_{l-}^{\ast }(k), \label{ty8}
\end{align}
\begin{align}
\begin{array}{c}
  \lim\sum_{l=-\infty}^{\infty}A_{22}(l,l) \\
  t\rightarrow\infty~~~~~~~~~~~~~~~~~~
\end{array}
=&\sum_{k}|\mathbf{a}_{k}|^{2} \sum_{l}|\mathbf{{b}}_{l+}(k)|^{2}\notag \\
+&\sum_{k}|\mathbf{b}_{k}|^{2} \sum_{l}| \mathbf{{b}}_{l-}(k)|^{2}. \label{ty10}
\end{align}
According to Eqs.(\ref{chirality}), (\ref{qdet}) and (\ref{asym}),
we define
\begin{equation}
\Pi_{L\pm}\equiv\sum_{l}| \mathbf{{a}}_{l\pm}(k)|^{2}, \label{ty66}
\end{equation}
\begin{equation}
\Pi_{R\pm}\equiv\sum_{l}|\mathbf{{b}}_{l\pm}(k)|^{2}, \label{ty77}
\end{equation}
\begin{equation}
Q_{0\pm}\equiv\sum_{l} \mathbf{{a}}_{l\pm}(k) \mathbf{
{b}}_{l\pm}^{\ast }(k). \label{ty88}
\end{equation}
To obtain the explicit shape of $\Pi_{L\pm}$, $\Pi_{R\pm}$ and
$Q_{0\pm}$, we can use again Eqs.(\ref{estacio}, \ref{q0entengl})
where the initial condition is, the walker localized at the position
$k$ with chirality $|\pm\rangle $ respectively, \emph{i.e.}
\begin{equation}
|\Psi_{2\pm} (0)\rangle = |\pm\rangle |k\rangle .  \label{psi1}
\end{equation}
Note that the principal difference between Eq.(\ref{psi0}) and
Eq.(\ref{psi1}) is in the walker's initial position, that is, in
Eq.(\ref{psi1}) $k$  is arbitrary and in Eq.(\ref{psi0}) $k=0$.
However, even with this difference, Eq.(\ref{qdet}) continues to be
valid for the calculation of $Q_{0\pm}$ because the original map,
Eq.(\ref{mapa}), is invariant under translations and therefore this
magnitude is independent of $k$. For the same reason
Eqs.(\ref{ty66}) and (\ref{ty77}) are independent of $k$, and then
Eq. (\ref{density300}) reduces to
\begin{equation}
\rho_{2c}=\Pi _{L}\rho_{L}+\Pi _{R}\rho_{R},  \label{redu2}
\end{equation}
where
\begin{align}
\rho_{L}&=\left(
\begin{array}{cc}
\Pi_{L+}& Q_{0+} \\
Q_{0+}^{\ast }&\Pi _{R+}
\end{array}%
\right),  \label{rhoL}
\end{align}
\begin{align}
\rho_{R}&=\left(
\begin{array}{cc}
\Pi_{L-}& Q_{0-} \\
Q_{0-}^{\ast }&\Pi _{R-}
\end{array}
\right),  \label{rhoR}
\end{align}
and $\Pi_{L}$ and $\Pi_{R}$ are given by Eqs.(\ref{estacio}, \ref{q0entengl}). Moreover, using the initial condition of
Eq.(\ref{psi1}) in Eq.(\ref{q0entengl}) and Eq.(\ref{estacio}) it is straightforward to obtain
\begin{align}
Q_{0+}=&-Q_{0-}=\frac{1}{2}-\frac{1}{2\sqrt{2}},  \notag \\
\Pi_{L+}=&~~\Pi_{R-}=1-\frac{1}{2\sqrt{2}},  \notag \\
\Pi_{R+}=&~~\Pi_{L-}=\frac{1}{2\sqrt{2}}. \label{q0+}
\end{align}
Note that $\rho_{L}$ and $\rho_{R}$, Eqs.(\ref{rhoL}, \ref{rhoR}),
are the asymptotic densities that correspond to initial conditions
associated with the eigenvalues $|+\rangle$ and $|-\rangle$
respectively. Therefore, Eq.(\ref{redu2}) has a natural
interpretation, after the measurement the asymptotic reduced density
matrix is the weighted average of the two densities associated with
the two possible values of the chirality.

\section{Entropy change}
The unitary evolution of the QW generates entanglement between the
coin and position degrees of freedom which can be quantified through
the associated von Neumann entropy. This entropy of entanglement is
defined by the reduced density operator
\begin{equation}
S(\rho_{c} )=-\kappa \ \mathrm{tr}(\rho_{c}\log \rho_{c}),
\label{dos}
\end{equation}
where $\kappa$ denotes a proportionality constant, the Boltzmann constant.

For a pure state the minimum entropy $S(\rho_{c} )=0$ is attained.
The maximum entropy (or minimum purity) is to be found for the
broadest possible probability distribution, the equipartition over
all pure states.

Equation (\ref{dos}) can be expressed as a function of the
eigenvalues of $\rho_{c}$, $\Lambda_{+}$ and $\Lambda_{-}$
\begin{equation}
\frac{S(\rho_{c} )}{\kappa}=-\Lambda_{+}\log \Lambda_{+}-\Lambda_{-}\log \Lambda_{-}. \label{doss}
\end{equation}

Using Eqs.(\ref{densityc}), (\ref{redu2}), (\ref{rhoL}) and
(\ref{rhoR}) we can calculate the entanglement entropies for the
four stationary densities introduced in the previous section. The eigenvalues of the density
operator $\rho_{1c}$ are
\begin{equation}
\Lambda _{1\pm}=\frac{1}{2}\pm \sqrt{2\mu^2+\nu^2}, \label{lam0}
\end{equation}
and those of $\rho_{2c}$ are
\begin{equation}
\Lambda _{2\pm}=\frac{1}{2}\pm |\mu|(\sqrt{2}-1). \label{lam22}
\end{equation}
The operators $\rho_{L}$ and $\rho_{R}$ have the same eigenvalues
and they are
\begin{align}
\Lambda_{LR\pm}=&\frac{1}{2}\mp\frac{1}{2}\sqrt{3-\frac{3}{\sqrt{2}}}.
\label{lamm33}
\end{align}
Therefore, the four entanglement entropies are expressed as
\begin{align}
\frac{S(\rho_{1c})}{\kappa}=&-\Lambda_{1+}\log \Lambda_{1+}-\Lambda_{1-}\log \Lambda_{1-},\label{ss1}\\
\frac{S(\rho_{2c})}{\kappa}=&-\Lambda_{2+}\log \Lambda_{2+}-\Lambda_{2-}\log \Lambda_{2-},\label{ss2}\\
\frac{S(\rho_{L})}{\kappa}=&-\Lambda_{LR+}\log \Lambda_{LR+}-\Lambda_{LR-}\log \Lambda_{LR-},\label{ss3}\\
\frac{S(\rho_{R})}{\kappa}=&~~\frac{S(\rho_{L})}{\kappa}\approx0.139~. \label{ss}
\end{align}

We now show that $S(\rho_{2c})$ has a non obvious upper bound, using a known
theorem that plays an important role in many applications of quantum information theory \cite{chuang2}.
Applying the mentioned theorem to the entropy of the mixture of quantum states given by
Eq. (\ref{redu2}), the following upper bound is obtained
\begin{equation}
S_2\equiv S(\rho_{L})-\kappa\left(\Pi _{L}\log \Pi _{L}+\Pi _{R}\log \Pi _{R}\right)\geq S(\rho_{2c}). \label{teorema2a}
\end{equation}
The entropies $S(\rho_{1c})$, $S(\rho_{2c})$ and the expression $S_2$ only depend
on the interference term $Q_{0}$.

Note that from Eq.(\ref{q0entengl}) we have $|Q_0|\leq
\frac{1}{2}(1-\frac{1}{\sqrt{2}})<1$, then we
approximate the $\log$ functions in Eqs. (\ref{ss1}), (\ref{ss2})
and (\ref{teorema2a}), using the first few terms of their
Taylor series to calculate the entropy change between the two asymptotic
stationary states,
\begin{equation}
\frac{S(\rho_{2c})-S(\rho_{1c})}{\kappa}\approx 2\nu^{2}+
2(2\sqrt{2}-1)\mu^2. \label{deltaS}
\end{equation}
The distance between $S(\rho_{2c})$ and its upper bound can be approximated by
\begin{equation}
\frac{S_2-S(\rho_{2c})}{\kappa}\approx
\frac{S(\rho_{L})}{\kappa}-4(\sqrt{2}-1)\mu^2. \label{deltaS2}
\end{equation}
Combining Eqs.(\ref{teorema2a}), (\ref{deltaS}) and (\ref{deltaS2}),
it is easy to obtain the following upper bound for
$S(\rho_{2c})-S(\rho_{1c})$
\begin{equation}
J_1\equiv S(\rho_{L})+2\kappa(\mu^2+\nu^2)\geq
S(\rho_{2c})-S(\rho_{1c}). \label{deltaS3}
\end{equation}

When $Q_{0}$ vanishes, Eq.(\ref{deltaS}) shows that $S(\rho_{1c})=S(\rho_{2c})$ and these entropies take their maximum
value. Moreover in this case ($Q_{0}=0$) the QW dynamics can be described as a classical Markovian process \cite{Alejo2};
it has a Markovian behavior both before and after the measurement. The initial conditions for this behavior are
$\gamma=-\pi/4, \pi/4$ and $\varphi=0,\pi$ respectively. When $Q_{0}\ne 0$, the measurement process essentially
determines an entropy increase for the system, $S(\rho_{2c})-S(\rho_{1c})>0$.

\begin{figure}[th]
\begin{center}
\includegraphics[scale=0.3]{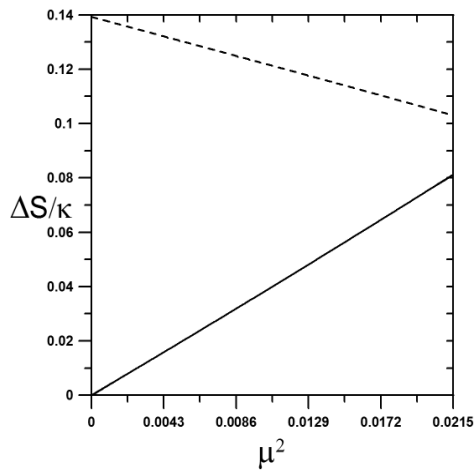}
\end{center}
\caption{Dimensionless entropy change as a function of
$|Q_0|^{2}=\mu^{2}$ for the case $\varphi=0$. In thick line the
entropy change given by Eq.(\ref{deltaS}). In dashed line the
entropy change given by Eq.(\ref{deltaS2}).} \label{f1}
\end{figure}

The entropy change for a real $Q_0$ is depicted in Fig.~\ref{f1}.
The calculation was made using the entropy definition Eqs.
(\ref{ss1}, \ref{ss2}, \ref{ss3}, \ref{ss}), therefore the curves
show that the lineal approximation proposed in Eqs. (\ref{deltaS},
\ref{deltaS2}) is excellent.

\section{Thermodynamic laws.}
In previous works \cite{Alejo2012,gustavo2014,renato2014,JCmodel},
we studied the behavior of the chirality distribution and we
introduced the temperature concept for an unitary closed
system. In this theory one considers
the system associated with the chirality degrees of freedom and
described by $\rho_c$, interacting through its entanglement
with the position degrees of freedom, the lattice, as equivalent to
the thermal contact with a bath.
Therefore, in equilibrium
\begin{equation}
[H_c,\rho_{c}]=0,  \label{scho2}
\end{equation}
should be satisfied, where $H_c$ is the interaction Hamiltonian
between the chirality and the lattice. In the QW case, the explicit
shape of $H_c$ is unknown for us, however we know that the
eigenvalues of $H_c$ are independent of the initial condition of the
wave function, they only depend on the unitary evolution. In
contrast, the eigenvalues $\Lambda _{\pm}$ depend on the
initial conditions and the corresponding eigenvalues of $H_c$. We
call $\{\left\vert\Phi_{\pm}\right\rangle\}$ the set of
eigenfunctions common to the density matrix and the Hamiltonian,
then in this basis the operators $\rho_{c}$ and $H_{c}$ are both
diagonal. Moreover, since only the relative difference between
energy eigenvalues has physical significance, we denote this set of
eigenvalues by ${\pm\epsilon}$; they may be interpreted as the
possible values of the entanglement energy. This interpretation
agrees with the fact that $\Lambda_{\pm}$ is the probability that
the system is in the eigenstate $\left\vert\Phi_{\pm}\right\rangle$.

The precise dependence between $\Lambda _{\pm}$ and $\pm\epsilon$ is
determined by the type of ensemble we construct. We propose that
our equilibrium state corresponds to a quantum canonical
ensemble. To this end we set
\begin{equation}
\Lambda _{\pm}\equiv\frac{e^{\mp\beta\epsilon}}{\mathcal{Z}},
\label{lam20}
\end{equation}
where $\mathcal{Z}$ is the partition function of the system, that is
\begin{equation}
{\mathcal{Z}}\equiv e^{-\beta\epsilon}+ e^{\beta\epsilon},
\label{part}
\end{equation}
and the parameter $\beta$ can be put into correspondence with an
entanglement temperature
\begin{equation}
T\equiv\frac{1}{\kappa\beta}=\frac{-2\epsilon}{\kappa\log\left({\Lambda _{+}}/{\Lambda _{-}}\right)}.  \label{tem}
\end{equation}
The entanglement temperature, Eq.(\ref{tem}), can take any finite, infinite, positive or negative value. Note that when
the energy of a system is bounded from above there is no compelling reason to exclude the possibility of negative
temperatures.

In this statistical mechanic frame it is possible to define the
internal energy of entanglement between the coin and position degrees
of freedom
\begin{equation}
U(\rho)=\epsilon\Lambda_{+}-\epsilon\Lambda_{-}. \label{uu2}
\end{equation}
The variations of
temperature, $\Delta T$ and internal energy, $\Delta U$ between the
two asymptotic states can be calculatedas as functions of the interference term $Q_0$.
Using Eqs.(\ref{lam0}) and (\ref{lam22}) together with Eqs.(\ref{tem})
and (\ref{uu2}) we have
\begin{align}
\frac{2\kappa\Delta T}{\epsilon}=&\frac{4}{\log\left({\Lambda
_{1+}}/{\Lambda _{1-}}\right)}-\frac{4}{\log\left({\Lambda
_{2+}}/{\Lambda _{2-}}\right)} \label{tem2} \\
\simeq&\frac{1}{\sqrt{2\mu^2+\nu^2}}
-\frac{1}{\left(\sqrt{2}-1\right)|\mu|}, \notag
\end{align}
\begin{equation}
\frac{\Delta U}{2\epsilon}=\left(\sqrt{2}-1\right)|\mu|-
\sqrt{2\mu^2+\nu^2}. \label{uu}
\end{equation}
Equation (\ref{tem2}) shows that $\Delta T=0$ is only possible if
$\Lambda _{1+}\Lambda _{2-}=\Lambda _{1-}\Lambda _{2+}$, and this
implies that $Q_0=0$, then the two asymptotic states
have the same temperature and in this sense we can to think of an
``isothermal process''. As the temperature concept makes sense only in an
equilibrium state it is clear that between these two
asymptotic states the temperature is not defined. Additionally
Eq.(\ref{uu}) implies that $\Delta U\leq0$ and vanishes when $Q_0=0$.

The first law of thermodynamics is now applied to the evolution between the
two asymptotic stationary states expressed as
\begin{equation}
\Delta U=\mathcal{Q}+\mathcal{W},  \label{prin}
\end{equation}
where $\mathcal{Q}$ is the QW heat absorbed during the measurement process and $\mathcal{W}$ is the work made over the
QW. Thermodynamic work is defined to be measurable solely from the knowledge of external constraint variables. In this
system the only parametric dependence of the thermodynamical functions is with the temperature, because the energy levels
$\{-\epsilon, \epsilon\}$ are maintained constant, then $\mathcal{W}=0$. Therefore the first law is reduced to
\begin{equation}
\Delta U=\mathcal{Q}.  \label{prin2}
\end{equation}
From Eq.(\ref{prin2}) we conclude that: i) The change of internal
energy is due to the heat delivered during the measurement process. ii)
The heat behaves as a state function, then the real process
can be substituted by a quasi-static process between the two
asymptotic stationary states (characterized by their temperatures),
where it is possible to defined the infinitesimal $d\mathcal{Q}$.
\begin{figure}[th]
\begin{center}
\includegraphics[scale=0.4]{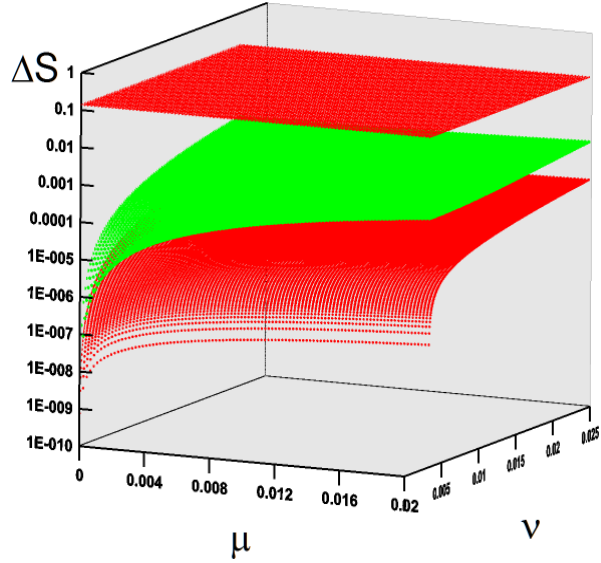}
\end{center}
\caption{(Color on line) The dimensionless entropy change (in $\log$ scale) as a function of the dimensionless
interference term $Q_0=\mu+i\nu$. The medium (green) surface is the change of entropy given by Eq.(\ref{deltaS}). The
upper and lower (red) surfaces are the upper and lower bounds $J_1$ and $J_2$ given by Eq.(\ref{deltaS3}) and
Eq.(\ref{prin33}) respectively. } \label{f2}
\end{figure}

In general, for an irreversible process, the second law of thermodynamics is expressed as
\begin{equation}
\Delta S>\int \frac{d\mathcal{Q}}{T}. \label{prin3}
\end{equation}
The question may be posed if this law is satisfied for the entropy, temperature and heat that were defined in our system. The
answer to this question must be positive because the starting point of our theory has been to postulate the
canonical distribution for the reduced density matrix
therefore the thermodynamic laws are obeyed. We do not know the temperature dependence with the
absorbed heat in order to compute the integral in Eq.(\ref{prin3}). However we can obtain a bound for this integral
in order to verify the second  law for the entanglement between the coin and position degrees of
freedom. We have $|T_1|<|T_2|$, where $T_1$ and $T_2$ are the temperatures of the stationary states before and after the measurement. Then
\begin{equation}
J_2\equiv\frac{\mathcal{Q}}{\ \ \ T_2}\leq \int
\frac{d\mathcal{Q}}{T}<S(\rho_{2c})-S(\rho_{1c}). \label{prin40}
\end{equation}
Therefore, using Eq.(\ref{prin2}) we can
propose the second law for the entanglement entropy
\begin{equation}
S(\rho_{2c})-S(\rho_{1c})>J_2=\frac{\Delta U}{\ \ \ T_2}\geq0.
\label{prin4}
\end{equation}
The lower bound $J_2$ can be expressed as a function of the
interference term $Q_0$, Eqs.(\ref{lam0}, \ref{lam22}, \ref{tem},
\ref{uu}, \ref{prin2}, \ref{prin4})
\begin{equation}
\frac{J_2}{\kappa}\simeq\left(\sqrt{2}-1\right)|\mu|\left[
\sqrt{2\mu^2+\nu^2} -\left(\sqrt{2}-1\right)|\mu|\right].
\label{prin33}
\end{equation}

In Fig.~\ref{f2} we show the entropy change together with its bounds. The entropy calculations  were made using the
definitions Eqs. (\ref{ss1}, \ref{ss2}). We conclude that the second law of thermodynamic in the form of Eq.(\ref{prin4})
is satisfied by the theory developed in the present paper and additionally we show the correctness of the upper bound
given by equation  Eq.(\ref{deltaS3}).
\section{Conclusion}
In previous works we developed the thermodynamics associated with the entanglement between the coin and
position degrees of freedom. Here we consider a special dynamics of a QW on a line. Initially, the walker localized at
the origin with arbitrary chirality, evolves to an asymptotic stationary state, then a measurement
is performed and the state resulting from this measurement is the initial condition for a second QW dynamics to achieve  a
second asymptotic stationary state.

We have studied the first and second laws of thermodynamics in the process between the two stationary states mentioned before.
These asymptotic stationary states only depends on the initial conditions through the interference term of the initial
wave function. We show that the change of entropy has upper and lower bounds and they are obtained analytically as a
function of the initial conditions. We have also shown that the measurement process changes the energy and this
change is associated to a heat transference process.

Moreover, we prove that, if the interference term vanishes the thermodynamics functions of the asymptotic stationary
states do not change.

\bigskip
I acknowledge stimulating discussions with V\'{\i}ctor Micenmacher, and the support from PEDECIBA (Programa de Desarrollo
de las Ciencias B\'asicas) and ANII (Agencia Nacional de Investigaci\'on e Innovaci\'on) (FCE-2-211-1-6281, Uruguay), and
the CAPES-UdelaR collaboration program.


\bigskip


\begin{thebibliography}{99}
\bibitem{Aharonov} Y. Aharonov, L. Davidovich, and N. Zagury, Phys. Rev. A \textbf{%
48}, 1687 (1993).
\bibitem{Meyer} D. A. Meyer, J. Stat. Phys. \textbf{85}, 551 (1996).
\bibitem{Watrous} J. Watrous, Proc. 33rd Symp. on the Theory of Computing (STOC'01) (ACM Press, New York, 2001), p.60.
\bibitem{Nayak} A. Nayak, and A. Vishwanath, arXiv:quant-ph/0010117 (2000).
\bibitem{Ambainis} A. Ambainis, Int.J. Quant. Inf. \textbf{1}, 507 (2003).
\bibitem{Kempe} J. Kempe, Contemp. Phys. \textbf{44}, 307 (2003).
\bibitem{Kendo1} V. Kendon, Math. Struct. Comp. Sci. \textbf{17}, 1169 (2006).
\bibitem{Kendo2} V. Kendon, Phil. Trans. R. Soc. A \textbf{364}, 3407 (2006).
\bibitem{Konno} N. Konno, \textit{Quantum Walks}, in Quantum Potential Theory, Lect. Notes Math., Vol. 1954, pp 309-452
ed. by U. Franz and M. Sch\"{u}rmann (Springer, 2008)
\bibitem{Shenvi} N. Shenvi, J. Kempe, K. BirgittaWhaley, Phys. Rev. A
\textbf{67}, 052307 (2003).
\bibitem{Ambainis2} A. Ambainis, SIAM Journal on Computing \textbf{37}, 210 (2007).
\bibitem{Childs0} A. M. Childs, R. Cleve, E. Deotto, E. Farhi, S. Gutmann, and D. A. Spielman, STOC Proc., pp. 59�68,
(2003), quant-ph/0209131.
\bibitem{Childs1}A. M. Childs and J. Goldstone, Phys. Rev. A 70, 022314 (2004). 12. A. Tulsi, Phys. Rev. A \textbf{78}, 012310
(2008).
\bibitem{Childs} A. M. Childs, Phys. Rev. Lett., \textbf{102}, 180501 (2009).
\bibitem{Lovett} N. B. Lovett, S. Cooper, M. Everitt, M. Trevers, and V. Kendon, Phys. Rev. A, \textbf{81}, 042330
(2010).
\bibitem{alejo2010} A. Romanelli, Phys. Rev. A \textbf{81}, 062349 (2010).
\bibitem{alejo2011} A. Romanelli, Physica A, \textbf{390}, 1209 (2011).
\bibitem{armando2011} A. P\'erez, and A. Romanelli, J. Comput. Theor. Nanosci.,\textbf{10}, 1 (2013).
\bibitem{Alejo2012} A. Romanelli, Phys. Rev. A, \textbf{85}, 012319 (2012).
\bibitem{gustavo2014} A. Romanelli, G. Segundo, Physica A, \textbf{393}, 646 (2014).
\bibitem{renato2014} A. Romanelli, R. Donangelo, R. Portugal, and F.
Marquezino \emph{Phys. Rev. A}, \textbf{90} 022329 (2014).
\bibitem{JCmodel} A. Romanelli, R. Donangelo, A. Vallejo \emph{forthcoming} (2015).
\bibitem{Sagawa1}  T. Sagawa and M. Ueda,  Phys. Rev. Lett. \textbf{100}, 080403 (2008).
\bibitem{Sagawa2}  T. Sagawa and M. Ueda,  Phys. Rev. Lett. \textbf{102}, 250602 (2009) and Erratum Phys. Rev. Lett. \textbf{106}, 189901 (2011).
\bibitem{Landauer} R. Landauer, IBM J. Res. Dev. \textbf{5}, 183 (1961); Science \textbf{272}, 1914 (1996).
\bibitem{nano} C. Jarzynski, Phys. Rev. Lett. \textbf{78}, 2690 (1997);
               G. E. Crooks, Phys. Rev. E \textbf{60}, 2721 (1999);
               S. Mukamel, Phys. Rev. Lett. \textbf{90}, 170604 (2003);
               R. Kawai et al., Phys. Rev. Lett. \textbf{98}, 080602 (2007);
               J. Liphardt et al., Science \textbf{296}, 1832 (2002);
               M. Collin et al., Nature (London) \textbf{437}, 231 (2005);
               A. B\'erut et al., Nature (London) \textbf{483}, 187 (2012).
\bibitem{Cohen} C. Cohen-Tannoudji, Rev. Mod. Phys. \textbf{70}, 707 (1998).
\bibitem{chuang} M. Nielssen and I. Chuang, \textit{Quantum Computation and Quantum Information}, Cambridge University Press, (2000)
\bibitem{Romanelli09} A. Romanelli, Phys. Rev. A \textbf{80}, 042332 (2009).
\bibitem{Alejo2} A. Romanelli, A.C. Sicardi Schifino, R. Siri, G. Abal, A.
Auyuanet, and R. Donangelo, Physica A, \textbf{338}, 395 (2004).
\bibitem{Alejo1} A. Romanelli, A.C. Sicardi Schifino, G. Abal, R. Siri, and
R. Donangelo, Phys. Lett. A \textbf{313}, 325 (2003).
\bibitem{Alejo4} A. Romanelli, R. Siri, G. Abal, A. Auyuanet, and R.
Donangelo, Physica A, \textbf{347}, 395 (2005).
\bibitem{Montero} M. Montero, Quantum Inf. Process., DOI:10.1007/s11128-014-0908-6
(2015).
\bibitem{chuang2} see Ref. \cite{chuang}  p. 518:
``Suppose $\rho=\Sigma_{i}p_{i}\rho_{i}$, where $p_{i}$ are
some set of probabilities, and the $\rho_{i}$ are density operators.
Then
\begin{equation}
S(\rho)\leq -\kappa\ \Sigma_{i}p_{i}\log
p_{i}+\Sigma_{i}p_{i}S(\rho_{i}), \label{teorema}
\end{equation}
with equality if and only if the states $\rho_{i}$ have support on orthogonal subspaces''.
\end{thebibliography}
\end{document}